\documentstyle[preprint,epsf,aps]{revtex}

\begin{document}
\hoffset-0.5cm

\preprint{hep-th/9710090}

\title{Three-Point Spectral Density in QED and the Ward Identity at
  Finite Temperature} 

\author{Hou Defu\cite{address1} and U. Heinz}

\address{
   Institut f\"ur Theoretische Physik, Universit\"at Regensburg,\\
   D-93040 Regensburg, Germany
}

\date{\today}

\maketitle

\begin{abstract}

We derive the spectral representations of QED 3-point functions and
then explicitly calculate the $3$-point spectral densities in hard
thermal loop approximation within the real time formalism. The Ward
identities obeyed by the retarded and advanced 2- and 3-point
functions are discussed. We compare our results with those for hot
QCD  .

\end{abstract} 



\section{Introduction}
\label{sec1}

Spectral densities are important quantities in finite temperature
field theory \cite{LvW87,K89}.  But they are not easy quantities to
evaluate perturbatively at nonzero temperature, especially for 
many-point spectral densities. The two-point spectral densities have
been widely studied and applied to QGP studies
\cite{blaizot,mak,bellac,J93,kob,hou96,ek}. In Ref.~\cite{kob} the
Cutkosky rules for calculating the imaginary parts of thermal Green
functions using the formalism of Thermo-Field Dynamics (TFD) was
presented. Recently these cutting rules were reexamined  within the
Closed Time Path (CTP) formalism \cite{chou,peterH} and given a simple
physical interpretation\cite{L96}. The imaginary time cutting rules
for calculating two-point spectral densities were investigated in
\cite{J93}. Three-point spectral densities, on the other hand, have so
far received less quantitative attention. The only available
calculation for gauge theories is published in \cite{taylor} where the
$3$-point spectral densities for pure gluon dynamics were calculated
in the Hard Thermal Loop (HTL) approximation using the imaginary time
formalism (ITF) \cite{pisarski}. The spectral representation of
three-point functions for selfinteracting scalar fields were 
discussed in Refs.~\cite{kob1,evan1,CH96,hh}.

Hard Thermal Loops (HTLs) are gauge invariant and satisfy simple
abelian Ward identities \cite{pisarski}. These remarkable properties
have triggered many interesting investigations
\cite{bp2,frank,wong,jac,Blz,kelly}. The computation oh HTLs is
generally fairly technical because of their complicated momentum and
energy dependence, but it can be simplified by using the Ward
identities \cite{pisarski,taylor1}. All HTLs can be derived from a
generating functional based on an effective Lagrangian
\cite{bp2}. They describe classical aspects of hot field theories 
and can thus also be obtained from classical kinetic equations
\cite{Heinz,Blz,kelly}. However, as Taylor stated in \cite{taylor},
for many purposes one does not need the HTL amplitudes themselves, but
only their discontinuities which are described by spectral
densities. These spectral densities also provide a natural connection
between the Green functions in the real time formulation (RTF) of
thermal field theory and the ITF. While the HTL resummation method was
developed within the imaginary time formalism (ITF), with primary  
attention focussed on equilibrium properties of hot field theories,
realistic physical systems are frequently out of thermal equilibrium
and require calculations in real time. Recently this has motivated
increased interest in the real time formulation of thermal field
theories and its non-equilibrium extensions.

In the present paper we therefore study the spectral functions and
Ward identities for finite temperature QED in the real time formalism
(RTF). Extending the recently derived, completely general spectral
representation of the real-time 3-point vertex function at finite
temperature from scalar field theory to the case of QED, we then show
how to implement the HTL approximation in real time. As shown in
\cite{CDT97}, for 2-point functions this procedure allows for a
simple generalization of the HTL resummation scheme to general
non-equilibrium situations. Within the HTL approximation we evaluate
explicitly the 3-point spectral densities and derive a set of finite
temperature Ward identities between the real-time 2- and 3-point
functions. Although one needs in general two independent spectral
densities to describe the real-time 3-point vertex at finite
temperature \cite{CH96,hh}, the two are shown to become degenerate in
the HTL approximation. This agrees with previous findings in
Refs.~\cite{taylor,Blz}. Our Ward identities between the real-time
Hard Thermal Loops in QED also agree with the general real-time finite
temperature Ward identities recently derived in Ref.~\cite{olivo97}. 

Throughout this paper we will use the CTP formalism \cite{chou} 
in the form given in Refs.~\cite{peterH,CH96}. In this representation of 
the real-time formalism the bosonic single-particle propagator in
momentum space has the form 
 \begin{equation}
 \label{1}
   D(p) = \left(  \matrix {D_{11} & D_{12} \cr
                           D_{21} & D_{22} \cr} \right) 
 \end{equation}
with
 \begin{mathletters}
 \label{2}
 \begin{eqnarray}
  i\, D_{11}(p) &=& \left(i\,D_{22}\right)^*
  = i{\cal P} \left(\frac {1}{p^2-m^2}\right) 
  + \left( n(p_0)+{1\over 2} \right) \rho(p)\, ,
 \label{2a} \\
  i\, D_{12}(p) &=&  n(p_0)\, \rho(p) \, ,
 \label{2b}\\
  i\, D_{21}(p) &=& 
  \bigl(1 + n(p_0) \bigr)\, \rho(p) \, .
 \label{2d}
 \end{eqnarray}
 \end{mathletters}
Here $n(p_0)$ is the thermal Bose-Einstein distribution,
 \begin{equation}
 \label{3}
   n(p_0) = \frac{1}{ e^{\beta p_0}-1},
 \end{equation}
and $\rho(p)$ is the two-point spectral density which for free particles
is given by
 \begin{equation}
 \label{3a}
   \rho(p)= 2\pi\, {\rm sgn}(p_0) \,\delta(p^2-m^2)\, .
 \end{equation}
The fermionic $2\times 2$ propagators are given by
 \begin{equation}
 \label{4}
   S_{ab}(p)=({\not\! p} +m)\tilde S_{ab}\, , \quad (a,b=1,2)\, ,
 \end{equation}
with
 \begin{mathletters}
 \label{5}
 \begin{eqnarray}
   i\, \tilde S_{11}(p) &=& 
   \left(i\,\tilde S_{22}\right)^*
   = i{\cal P} \left(\frac {1}{p^2-m^2}\right) 
     + \left(- \tilde n(p_0)+{1\over 2} \right) \rho(p)\, ,
 \label{5a} \\
   i\, \tilde S_{12}(p) &=& - \tilde n(p_0)\, \rho(p) \, ,
 \label{5b}\\
   i\, \tilde S_{21}(p) &=& 
   \bigl(1 - \tilde n(p_0) \bigr)\, \rho(p) \, .
 \label{5d}
 \end{eqnarray}
 \end{mathletters}
Here $\tilde n(p_0)$ is the thermal Fermi-Dirac distribution,
 \begin{equation}
 \label{6}
   \tilde n(p_0) = \frac{1}{ e^{\beta p_0}+1}\, ,
 \end{equation}
and for free fermions $\rho(p)$ is again given by (\ref{3a}).

The paper is organized as follows. In Sec.~\ref{sec2} we derive the
spectral representations for the retarded 3-point functions in QED. 
In Sec.~\ref{sec3} we evaluate the 3-point spectral densities for QED
in the HTL approximation. In Sec.~\ref{sec4} we derive the RTF Ward
identities between the 2- and 3-point HTL amplitudes in QED. A short
summary is given in Sec.~\ref{sec5}.

\section{Spectral representation of the 3-point vertex in QED}
\label{sec2}

In this Section we shortly review some useful relations among the
different thermal components of the QED 3-point functions and derive
their spectral representation. Similar  relations for the 3-point
vertex in $\phi^3$ theory  have been reported in the literature
\cite{kob,kob1,evan1} in different notation. Our procedure here
follows the notation developed in \cite{hh} for $\phi^3$ theory.
We consider the 3-point vertex function in QED shown in Fig.~1. 
The three incoming external momenta are $k_1=p$, $k_2=q$, and $k_3=-p-
q$.  

\begin{eqnarray}
\parbox{14cm}
{{
\begin{center}
\parbox{10cm}
{
\epsfxsize=7cm
\epsfysize=6cm
\epsfbox{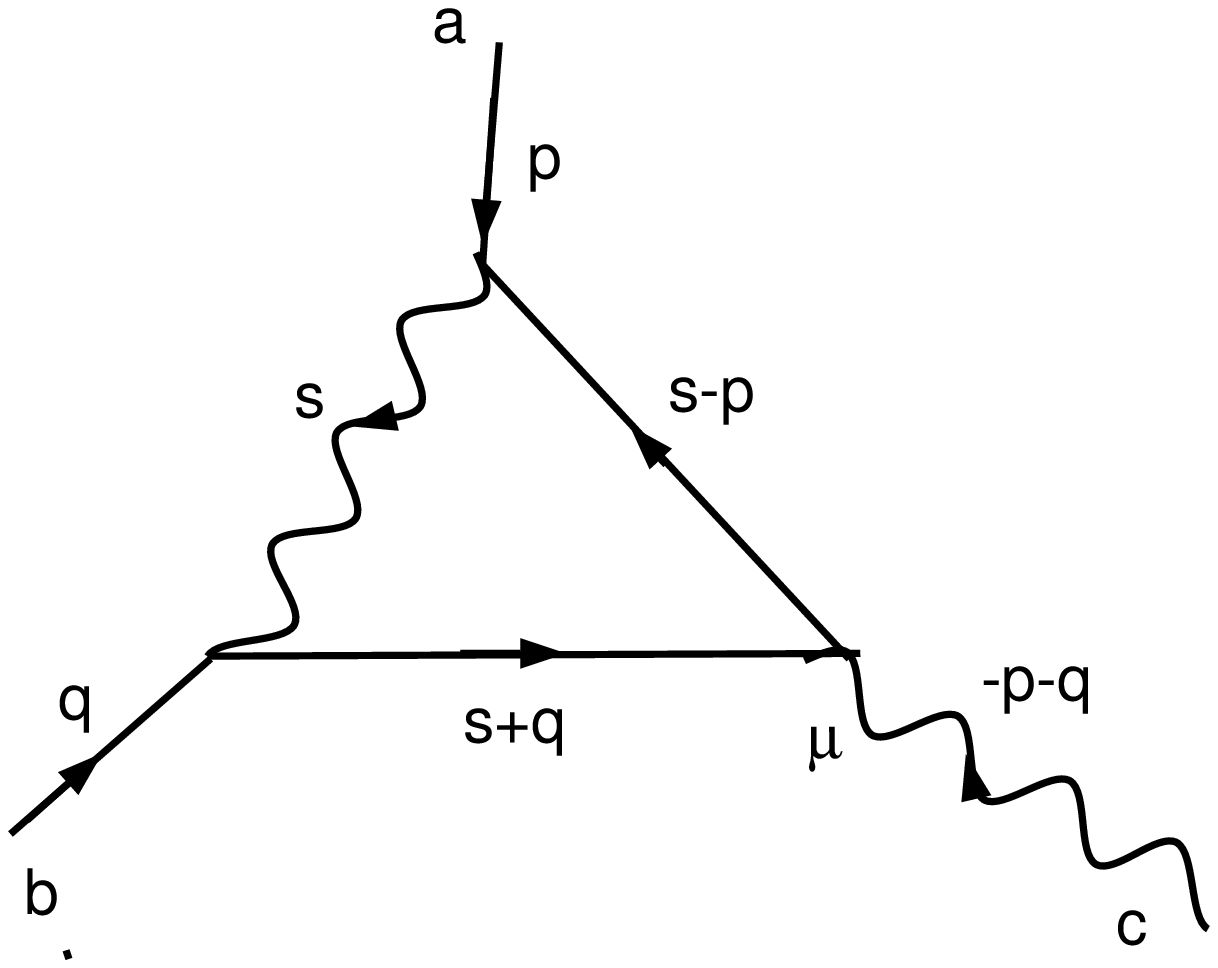}}\\
\parbox{14cm}{\small \center  Fig.~1: 3-point vertex in QED}
\end{center}
}}
\nonumber
\end{eqnarray}
Within the real time formalism, the truncated three-point function for
QED, $G^\mu_{abc}(x,y,z)$, has eight thermal components which satisfy
\cite{kob,chou}
 \begin{equation}
   \sum_{a,b,c=1}^2 G^\mu_{abc}=0\, .
 \label{c2}
 \end{equation}
Using the KMS condition one finds in momentum space\cite{kob2}
\begin{mathletters}
 \label{c3}
 \begin{eqnarray}
 \label{c3a}
   \tilde G^\mu_{111}(k_1,k_2,k_3) &=& - G^{\mu *}_{111}(k_1,k_2,k_3)
   = G^\mu_{222} (k_1,k_2,k_3) \, ,
 \\
 \label{c3b}
   \tilde G^\mu_{121}(k_1,k_2,k_3) &=& - G^{\mu *}_{121}(k_1,k_2,k_3)
   = -e^{\beta \omega_2}\,G^\mu_{212}(k_1,k_2,k_3) \, ,
 \\
 \label{c3c}
   \tilde G^\mu_{211}(k_1,k_2,k_3) &=&- G^{\mu *}_{211}(k_1,k_2,k_3)
   = -e^{\beta \omega_1}\,G^\mu_{122}(k_1,k_2,k_3) \, ,
 \\
 \label{c3d}
   \tilde G^\mu_{112}(k_1,k_2,k_3) &=& -G^{\mu *}{112}(k_1,k_2,k_3) 
   = e^{\beta\omega_3}\, G^\mu_{221}(k_1,k_2,k_3)
 \, ,
 \end{eqnarray}
 \end{mathletters} 
where $\tilde G^\mu$ represents ``tilde conjugation'' of $G^\mu$ (see
\cite{hh}). Note that in Eqs.~(\ref{c3b},\ref{c3c}) the last equation
involves an additional minus sign relative to the scalar case
\cite{hh}, due to the fermionic legs. 

One can construct ``retarded'' vertex functions from the above eight
thermal components according to
 \begin{mathletters}
 \label{c1}
 \begin{eqnarray}
 \label{c1a}
  G^\mu_{R} &=& G^\mu_{111}+G^\mu_{112}+G^\mu_{211}+G^\mu_{212},
 \\
 \label{c1b}
  G^\mu_{Ri} &=& G^\mu_{111}+G^\mu_{112}+G^\mu_{121}+G^\mu_{122},
 \\
 \label{c1c}
  G^\mu_{Ro} &=& G^\mu_{111}+G^\mu_{121}+G^\mu_{211}+G^\mu_{221},
 \end{eqnarray}
 \end{mathletters}
where $G^\mu_{Ri}$ is the vertex function which in coordinate space is
retarded with respect to $x_0$, $G^\mu_{Ro}$ is retarded with respect
to $z_0$, and $G^\mu_R$ is retarded with respect to $y_0$. The legs
linked to $x,y, z$ correspond to an electron, electron, and photon,
respectively, and the corresponding distribution functions are $\tilde
n_1 = \tilde n(\omega_1),\, \tilde n_2= \tilde n(\omega_2),$ and
$n_3=n(\omega_3)$. Inversion of Eqs.~(\ref{c1}) with the help of
(\ref{c2}) (\ref{c3}) yields expressions for the thermal components
$G^\mu_{abc}$ in terms the above retarded functions:
 \begin{eqnarray}
  G^\mu(k_1,k_2,k_3) &=& G_R^\mu {\tilde n_1\choose 1-\tilde n_1} 
  {1\choose -1} {-n_3\choose 1+n_3} 
                        -\frac{1}{2}G^{\mu*}_R(N_1+N_3)
              {1\choose -1}{\tilde n_2\choose 1-\tilde n_2}{1\choose -1}
 \nonumber\\
          &+&G_{Ri}^\mu {1\choose -1} {\tilde n_2 \choose 1- \tilde n_2}
          {-n_3\choose 1+n_3} 
            - \frac{1}{2}G_{Ri}^{\mu*} (N_2+N_3)
              {\tilde n_1 \choose 1- \tilde n_1}{1\choose -1}{1\choose -1}
 \nonumber\\
          &+& G_{Ro}^\mu {\tilde n_1\choose 1- \tilde n_1} 
             {\tilde n_2 \choose 1- \tilde n_2} {1\choose -1} 
             -\frac{1}{2}G_{Ro}^{\mu*} (N_1+N_2)
              {1\choose -1}{1\choose -1}{-n_3 \choose 1+n_3},
 \label{eq: DECOMP}
 \end{eqnarray}
with $N_1=1-2\tilde n_1$, $N_2=1-2\tilde n_2$, and $ N_3 = 1 + 2 n_3$.
The structure of this equation is similar to Eq.~(35) in
Ref.~\cite{CH96}; the sign differences arise from additional minus
signs in front of the Fermi distributions and from the fact that
Eq.~(\ref{eq: DECOMP}) refers to the truncated vertex rather than
the connected vertex studied in \cite{CH96}.

Following the same procedure as in Appendix A2 of Ref.~\cite{hh}
one derives the following spectral integral representation for the
retarded 3-point functions in QED:
 \begin{mathletters}
 \label{c21}
 \begin{eqnarray}
 \label{c21a}
   G^\mu_{R}(\omega_1,\omega_2,\omega_3) &=&
   \frac{-i}{2\pi^2} \int_{-\infty}^{\infty}
   \frac{d\Omega_1 d\Omega_2}{\omega_2-\Omega_2+i\epsilon}
   \left(
   \frac{\rho^\mu_1}
        {\omega_1-\Omega_1-i\epsilon}
  +\frac{\rho^\mu_1-\rho^\mu_2}
        {\omega_3-\Omega_3-i\epsilon} \right) \, ,
 \\
 \label{c21b}
   G^\mu_{Ri}(\omega_1,\omega_2,\omega_3) &=&
   \frac{-i}{2\pi^2} \int_{-\infty}^{\infty}
   \frac{d\Omega_1 d\Omega_2}{\omega_1-\Omega_1+i\epsilon}
   \left(
   \frac{\rho^\mu_2}
        {\omega_2-\Omega_2-i\epsilon} 
  -\frac{\rho^\mu_1-\rho^\mu_2}
        {\omega_3-\Omega_3-i\epsilon} \right)\, ,
 \\
 \label{c21c}
   G^\mu_{Ro}(\omega_1,\omega_2,\omega_3) &=&
   \frac{-i}{2\pi^2}\int_{-\infty}^{\infty}
   \frac{d\Omega_1 d\Omega_2}{\omega_3-\Omega_3+i\epsilon}
   \left(
   \frac{\rho^\mu_1}
        {\omega_1-\Omega_1-i\epsilon} 
  +\frac{\rho^\mu_2}
        {\omega_2-\Omega_2-i\epsilon} \right)\,.
 \end{eqnarray}
 \end{mathletters}
Here $\omega_1+\omega_2+\omega_3=\Omega_1+\Omega_2+\Omega_3=0$ and 
the spatial momenta $\bbox{p}_1$, $\bbox{p}_2$,
$\bbox{p}_3{=}-(\bbox{p}_1{-}\bbox{p}_2)$ are the same on both sides
and have therefore been suppressed. The spectral densities are given
by the following thermal components of the 3-point vertex in momentum
space: 
 \begin{mathletters}
 \label{A19}
 \begin{eqnarray}
 \label{A19a}
   \rho^\mu_1 &=& {\rm Im\, } (G^\mu_{122} - G^\mu_{211}) \, ,
 \\
 \label{A19b}
   \rho^\mu_2 &=& {\rm Im\, } (G^\mu_{212} - G^\mu_{121}) \, .
 \end{eqnarray}
 \end{mathletters}
One notes that the spectral representations for the truncated QED
three-point functions have the same form as in scalar $\phi^3$
theory \cite{hh}, except for the additional vector index. 

\section{Evaluation of the spectral densities in HTL approximation} 
\label{sec3}

In this Section we calculate the three-point  spectral functions
$\rho^\mu_1$, $\rho^\mu_2$ for QED in the hard thermal loop
approximation \cite{pisarski}. Since HTLs are gauge invariant we can
choose Feynman gauge for simplicity.  

From Eqs.~(\ref{c3}) and (\ref{A19}) we have
 \begin{equation}
 \label{15}
   \rho^\mu_1 = {\rm Im\,} (G_{122}^\mu - e^{\beta p_0}G_{122}^{*\mu})
   = {1\over \tilde n(p_0)} {\rm Im\, } G_{122}^\mu \, .
 \end{equation}
For $\rho^\mu_1$ we thus must evaluate only the single Feynman diagram 
in Fig.~1 for $a=1$, $b=c=2$. Using standard real-time Feynman rules
\cite{chou} and the photon propagator in Feynman gauge one gets 
 \begin{equation}
 \label{16}
   G_{122}^\mu(p,q,-p-q)=(-ig)(ig)^2\int \frac {d^4 s}{(2\pi)^4} \,
   [iD_{12}(s)]\gamma_\alpha [iS_{22}(s+q)]\gamma^\mu
[iS_{21}(s-p)]\gamma^\alpha\, .
 \end{equation}
Inserting the thermal free propagators (\ref{2})-(\ref{6}) and extracting the 
imaginary part one finds
 \begin{eqnarray}
   {\rm Im\,}G_{122}^\mu&=&-g^3 \int \frac {d^4 s}{(2\pi)^4}\,
   [\gamma_\alpha ({\not \!s} - {\not\!p} + m) 
    \gamma^\mu ({\not\!s} + {\not\! q} + m)\gamma^\alpha] \,
 \nonumber\\
   && \quad \times \,
   \delta(s^2) \, \delta((s+q)^2-m^2) \, \delta((s-p)^2-m^2) \,
   {\rm sgn}(s_0)\, {\rm sgn}(s_0+q_0)\, {\rm sgn}(s_0-p_0)
\nonumber\\
   && \quad \times \,
      n(s_0) \, \Bigl( {\textstyle{1\over 2}} - \tilde n(s_0+q_0) \Bigr) 
      \, \bigl( 1 - \tilde n(s_0-p_0) \bigr)
\end{eqnarray}

If the coupling constant $g$ is small and the external momenta  are
soft, $p,q\sim gT$, and the electron bare mass $m$ is much smaller
than the temperature, the leading contributions comes from the hard
loop momenta $s\sim T$ \cite{mak,pisarski}. For these  we can thus
neglect the external momenta and the mass $m$ in the terms between
square brackets, approximating them by $\gamma_\alpha {\not\! s}
\gamma_\mu {\not\! s} \gamma_\alpha = - 4 s_\mu {\not\! s} +2
\gamma_\mu s^2$. After performing the integration over $s^0$ with the
help of the function $\delta(s^2) = [\delta(s_0-\bar s) +
\delta(s_0+\bar s)]/2 \bar s$, where $\bar s = \sqrt{{\bbox{s}^2}}$,
one finds 
 \begin{mathletters}
 \label{17}
 \begin{eqnarray}
 \label{17a}
   \rho_1^\mu(p,q,-p-q) &=& 
   {g^3 \over n(p_0)} \Bigl( A^\mu(p,q) + B^\mu(p,q) \Bigr)\, ,
 \\
 \label{17b}
   A^\mu(p,q) &=&\int {d^3 s\over (2\pi)} \, {1\over 2 \bar s}
          {\rm sgn}(\bar s+q_0)\, {\rm sgn}(\bar s-p_0 ) 
   4s^\mu{\not\! s} \vert_{s^0=\bar s}
 \nonumber\\
   && \quad \times 
      \delta\bigl( (\bar s+q_0)^2 - E_{s+q}^2 \bigr) \,
      \delta\bigl( (\bar s-p_0)^2 - E_{s-p}^2 \bigr) 
 \nonumber\\
   && \quad \times 
      n(E_s) \, \Bigl( {\textstyle{1\over 2}} - \tilde n(\bar s+q_0) \Bigr) 
      \, \bigl( 1 - \tilde n(\bar s-p_0) \bigr)\, ,
 \\
 \label{17c}
   B^\mu(p,q) &=& -\int {d^{n-1}s\over (2\pi)^{n-3}} \, {1\over 2 \bar s}
          {\rm sgn}(\bar s-q_0)\, {\rm sgn}(\bar s+p_0 )
   4s^\mu{\not\! s} \vert_{s^0=- \bar s}
 \nonumber\\
   && \quad \times 
      \delta\bigl( (\bar s-q_0)^2 - E_{s+q}^2 \bigr) \,
      \delta\bigl( (\bar s+p_0)^2 - E_{s-p}^2 \bigr) 
 \nonumber\\
   && \quad \times 
      \bigl( n(-\bar s) \bigr) \, 
      \Bigl( {\textstyle{1\over 2}} - \tilde n(-\bar s+q_0) \Bigr) \, 
      \tilde n(\bar s+p_0)\, .
 \end{eqnarray}
 \end{mathletters}
Here $E_{s+q} = \sqrt{m^2 + (\bbox{s}+\bbox{q})^2}$, $E_{s-p} = 
\sqrt{m^2 + (\bbox{s}-\bbox{p})^2}$, and in (\ref{17c}) we used the 
identity $n(-x)+n(x) = -\eta$, $\eta=\pm 1$ for bosons and fermions. 

To simplify the notation it is convenient to introduce the 4-vectors
$V_\pm = \left(1, \pm {\bbox{s}\over \bar s} \right)$. For hard loop
momenta the arguments of the $\delta$-functions in (\ref{17}) can then
be written as 
\begin{mathletters}
 \label{20}
 \begin{eqnarray}
 \label{20a}
    \left. (s+q)^2 \right\vert_{s^0=\pm \bar s} 
    &\approx& \pm 2\bar s\, q\cdot V_\pm 
     = \pm 2\bar s\left(q^0 \mp |\bbox{q}| \cos\theta'\right)\, ,
 \\     
 \label{20b}
    \left. (s-p)^2 \right\vert_{s^0=\pm \bar s} 
    &\approx& \mp 2\bar s\, p\cdot V_\pm 
     = \mp 2\bar s \left(p^0 \mp |\bbox{p}| \cos\theta\right)\, ,
 \end{eqnarray}
 \end{mathletters}
and we can replace ${\rm sgn}(\bar s \pm p^0) \approx 1 \approx {\rm
  sgn}( \bar s \pm q^0)$ under the integral. $\theta$ and $\theta'$
are the angles between $\bbox{s}$ and $\bbox{p}, \bbox{q}$, respectively.
These approximations decouple the angular and radial integrations
\cite{pisarski}; we obtain
 \begin{mathletters}
 \label{22}
 \begin{eqnarray}
 \label{22a}
   A^\mu(p,q) &\approx& a(p_0) \, \omega^\mu(p,q)\, ,
 \\
 \label{22a1}  
   B^\mu(p,q) &\approx& b(p_0) \, \omega^\mu(p,q)\, ,
 \\
 \label{22b}
   a(p_0) &=& {1 \over 2 \pi} \int_0^\infty 
   d\bar s \, {4\bar s^4 \over (2\bar s)^3}  \, n(\bar s) \, 
   \Bigl( {\textstyle{1\over 2}} - \tilde n(\bar s+q_0 ) \Bigr) 
   \bigl(1 -  \tilde n(\bar s-p_0)\bigr) \, ,
 \\
 \label{22c}
   b(p_0) &=&-{1\over 2\pi} \int_0^\infty 
   d\bar s \, {4\bar s^4 \over (2\bar s)^3}  \, n(-\bar s) \, 
   \Bigl( {\textstyle{1\over 2}} - \tilde n(-\bar s+q_0 ) \Bigr) 
    \tilde n(\bar s+p_0)\bigr) \, ,
 \\
\label{22d}
   \omega^\mu (p,q) &=&\int d\Omega \,V_+^\mu {\not\! V}_+ \,
   \delta(q \cdot V_+) \, \delta(p \cdot V_+) =
   \int d\Omega \, V_-^\mu {\not\! V}_-\,\delta(q \cdot V_-) \, 
   \delta(p \cdot V_-) \, .
 \end{eqnarray}
 \end{mathletters}
Except for the additional spinor structure the angular integral
(\ref{22d}) is identical with the one found by Taylor \cite{taylor}
for the spectral density of the 3-gluon vertex in hot QCD. 

The integrands in Eqs.~(\ref{22b},\ref{22c}) contain up to three
powers of thermal distribution functions. By using following identities
 \begin{mathletters}
 \label{19}
 \begin{eqnarray}
 \label{19a}
   n(\omega_1) \tilde n(\omega_2)&=&\tilde n(\omega_1+\omega_2)
   (1+n(\omega_1)-\tilde n(\omega_2))\, ,
 \\
 \label{19b} 
   \tilde n(\omega_1) \tilde n(\omega_2)&=&n(\omega_1+\omega_2)(1-\tilde
   n(\omega_1)-\tilde n(\omega_2)), 
 \end{eqnarray}
 \end{mathletters}
one shows that the cubic terms disappear and that $a$ and $b$ reduce
to 
 \begin{mathletters}
 \label{23}
 \begin{eqnarray}
 \label{23a}
   a(p_0) &=& {1 \over 4 \pi} \int_0^\infty  d\bar s \, {\bar s} 
   \Bigl({\textstyle{1\over 2}} \tilde n(p_0)
   \left[n(\bar s)+\tilde n(\bar s-p_0)\right] \,
 \nonumber\\
   && - n(p_0+q_0) \Bigl[\tilde n(-q_0)[n(\bar s)+\tilde n(\bar s+q_0)]
      -\tilde n(p_0)[n(\bar s) +\tilde n(\bar s-p_0)]\Bigr]\Bigr) \, ,
 \\ 
 \label{23b}
   b(p_0) &=& {1 \over 4 \pi} \int_0^\infty  d\bar s \, {\bar s} 
   \Bigl({\textstyle{1\over 2}} \tilde n(p_0)
   \left[n(\bar s)+\tilde n(\bar s+p_0)\right] \,
 \nonumber\\
   && - n(p_0+q_0) \Bigl[\tilde n(-q_0)[n(\bar s)+\tilde n(\bar s-q_0)]
      -\tilde n(p_0)[n(\bar s) +\tilde n(\bar s+p_0)]\Bigr]\Bigr) \, ,
 \end{eqnarray}
 \end{mathletters}
The remaining integrands are linear in the $\bar s$-dependent thermal
distribution functions \cite{baier,hh}.

After evaluating the integral over $\bar s$ in the limit $p_0/T \ll
1,\, q_0/T \ll 1$ we obtain to leading order in the coupling constant
$g$ 
\begin{eqnarray}
 \label{24}
    \rho_1^\mu(p,q)&=&{g^3\over \tilde n(p_0)}
    \left(A^\mu(p,q)+B^\mu(p,q)\right) 
 \nonumber\\
    &\approx& {g m_{\beta}^2 \pi\over 2} 
    \int d\Omega \,V_+^\mu {\not\! V}_+ \, 
    \delta(q \cdot V_+) \, \delta(p \cdot V_+) 
\nonumber\\
    &=&{g m_{\beta}^2 \pi\over 2}
    \int d\Omega \, V_-^\mu {\not\! V}_- \, 
    \delta(q \cdot V_-) \, \delta(p \cdot V_-) \, .
 \end{eqnarray}
Here $m_{\beta}={gT\over \sqrt{8}}$ is the thermal electron mass.

The other spectral density $\rho_2$ is determined from
 \begin{equation}
 \label{25}
   \rho_2^\mu = {1\over \tilde n(q_0)} {\rm Im\ }G_{212}^\mu \, .
 \end{equation}
By inspection of the corresponding labelling of the diagram in Fig.~1
one observes that $G_{212}(p,q,-p-q)$ is obtained from
$G_{122}(p,q,-p-q)$ by exchanging the two electron legs with the
external momenta $p$ and $q$ and routing the internal momentum $s$ in
the opposite direction. The resulting loop integral then becomes
identical to the one before, Eq.~(\ref{16}), and we find
 \begin{mathletters}
 \label{26}
 \begin{eqnarray}
 \label{26a}
    \rho_2^\mu(p,q)&\approx &\rho_1^\mu(q,p)\approx \rho_{_{\rm
      HTL}}^\mu(p,q) \, , 
 \\
 \label{26b} 
    \rho_{_{\rm HTL}}^\mu(p,q) &\approx& {gm_{\beta}^2 \pi\over 2}  
    \int d\Omega\, V_+^\mu \,{\not\! V}_+ \, 
    \delta(p{\cdot}V_+)\, \delta(q{\cdot}V_+)\, .
 \end{eqnarray}
 \end{mathletters}
This agrees with the observation by Taylor within \cite{taylor} that in QCD
in HTL approximation the two independent spectral densities for
the 3-gluon vertex degenerate.

Considering the $\delta$-functions in Eq.~(\ref{26}) one easily shows
that
 \begin{equation}
  p_\mu \rho_{_{\rm HTL}}^\mu(p,q) = 0 = 
  q_\mu \rho_{_{\rm HTL}}^\mu(p,q)\, 
\end{equation}
i.e. in HTL approximation the QED spectral density is transverse with
respect to the external momenta. This nice feature was also noted in
\cite{taylor} for hot QCD. 

\section{Real-time Ward indentities among Hard Thermal Loops}
\label{sec4}

Let us insert the above spectral densities into the spectral
representations for the retarded 3-point functions:
 \begin{mathletters}
 \label{12}
 \begin{eqnarray}
 \label{12a}
   G_{R}^\mu(\omega_1,\omega_2,\omega_3) &=&
   -ig \, {m_\beta^2\over 4\pi} \int d\Omega
   {V_+^\mu {\not\! V}_+\over 
   (V_+\cdot k_1-i\epsilon)(V_+\cdot k_2+i\epsilon)} \, ,
 \\
 \label{12b}
   G_{Ri}^\mu(\omega_1,\omega_2,\omega_3) &=&
   -ig \, {m_\beta^2\over 4\pi} \int d\Omega
   {V_+^\mu {\not\! V}_+\over 
    (V_+\cdot k_1+i\epsilon)(V_+\cdot k_2-i\epsilon)} \, ,
 \\
 \label{12c}
   G_{Ro}^\mu(\omega_1,\omega_2,\omega_3) &=&
   ig \, {m_\beta^2\over 4\pi} \int d\Omega
   {V_+^\mu {\not\! V}_+\over 
   (V_+\cdot k_1-i\epsilon)(V_+\cdot k_2-i\epsilon)} \, .
 \end{eqnarray}
 \end{mathletters}
The QCD-analogue of last two expressions were previously obtained 
by Blaizot and Jancu \cite{Blz} from a set of classical kinetic
equations. If the external momenta are of order $gT$, $p_1\sim p_2\sim
gT$, power counting reveals that all three retarded 3-point functions
are of order $g$, i.e. of the same order as tree vertex. The HTLs
(\ref{12}) for the QED 3-point vertex in the real time formalism thus
require resummation within perturbation theory, like the corresponding
HTLs in the imaginary time formalism \cite{pisarski,bellac}. 

In the next step we derive the real-time analogue of the well-known
Ward identities between the HTL amplitudes in the ITF formalism
\cite{taylor1,bp2}. Due to the matrix structure of the real-time
thermal Green functions, the Ward identities also become matrix
equations.

The zero temperature Ward identities for QED can be written both in
differential and in integrated form \cite{Ward}:
 \begin{equation}
 \label{wd2}
   \frac{\partial S (p)}{\partial p^\mu}=
   \frac{i}{g} S(p)\, \Gamma_\mu(p,-p,0) \, S(p) \, ,
\end{equation}
or
 \begin{equation}
 \label{wd0}
  g\, \left(S(p')-S(p)\right) = i\,S (p) \left[
    (p'-p)^\mu \Gamma_\mu(p,-p',p'-p) \right] S(p') \, .
 \end{equation}
Multiplying by $S^{-1}(p)$ from the left and by $S^{-1}(p')$ from the
right and using the Schwinger-Dyson equation
 \begin{equation}
 \label{sw}
   S^{-1}(p) = {\not\! p} - m - \Sigma(p)\, ,
 \end{equation}
the Ward identity takes the form
 \begin{equation}
 \label{wdT1}
  (p'-p)^\mu  G_\mu(p,-p',p'-p) = -i g \,  [\Sigma(p')-\Sigma(p)] \, ,
\end{equation}
where $G_\mu$ is the vertex correction from loop diagrams, defined by
$G_\mu = \Gamma_\mu - \Gamma_\mu^0$, with $\Gamma_\mu^0 = i g \gamma_\mu$. 

At finite temperature the propagators and vertices have matrix
structure. The generalization of (\ref{wdT1}) to finite temperature in
the real time formalism was recently performed in Ref.~\cite{olivo97}.
The authors of this paper used a different version of the real time
formalism and considered the vertex with the photon attached to the
ingoing leg; translated into the CTP framework and for the vertex with
the photon on the outgoing leg their result reads:
 \begin{equation}
 \label{olivo}
  (p'-p)_\mu  G^\mu_{abc}(p,-p',p'-p) = -i g \,  
  [\delta_{bc}\Sigma_{ac}(p') - \delta_{ac}\Sigma_{cb}(p)] \, , 
 \end{equation}
where $a,b,c=1,2$. This can be rewritten in terms of the retarded
and advanced amplitudes; the result agrees with what we find for the
HTL amplitudes below.  

In order to examine the relations among the 2- and 3-point HTLs in the
real time formalism, we calculate the retarded electron self-energy 
\cite{peterH}: 
 \begin{eqnarray}
 \label{rs}
   \Sigma_R(p) = g^2 \int \frac{d^3\bbox{k} }{(2\pi)^3}
   \int d\omega\, d\omega'\, \rho_f(\omega, \bbox{p}+\bbox{k})\, 
   \rho_B(\omega', \bbox{k}) 
   \left(\frac{n(\omega')+\tilde n(\omega)}
              {p_0+\omega'-\omega+i\epsilon}\right) \, ,
 \end{eqnarray}
where $\rho_F$ and $\rho_B$ are the two-point spectral densities of the
fermions and bosons, respectively. Inserting the free particle
spectral densities
 \begin{eqnarray}
 \label{2pspec}
   \rho_B(\omega', \bbox {k})&=& {\rm sgn}(\omega')\,
   \delta(\omega'^2-E_k^2)\, ,
 \nonumber\\
   \rho_F(\omega, \bbox {p+k})&=&
   (\omega\gamma^0+\bbox{k'}\cdot\bbox{\gamma}+m) \,
   {\rm sgn}(\omega)\, \delta(\omega^2-E_{p+k}^2)\, ,
 \end{eqnarray}
with $E_k = \sqrt{\bbox {k}^2}$, $E_k'=\sqrt{\bbox {k'}^2+m^2}$,
$k'=k+p$, and evaluating the integral at in the HTL approximation 
and using dimensional regularization \cite{weldon}, we obtain
 \begin{eqnarray}
 \label{rs1}
   \Sigma_R(p)=a(p) {\not\! p} + b(p) \gamma_0 \, ,
 \end{eqnarray}
where
 \begin{eqnarray}
 \label{ab}
   a(p_0,\bbox{p}) &=& \frac{m_\beta^2}{\bbox{p}^2}
   \left(1 - \frac{p_0}{4\pi} \int d\Omega \, 
   \frac{1}{V_+\cdot p+i\epsilon}\right) \, ,
 \nonumber\\
   b(p_0,\bbox{p}) &=& \frac{m_\beta^2}{\bbox{p}^2} \left( p_0 +
   {p^2 \over 4\pi} \int d\Omega \, 
   \frac{1}{V_+\cdot p+i\epsilon}\right) \, .
\end{eqnarray}
Here $V_+=(1, \bbox{V})$,$ \bbox{V}= \frac{\bbox{k}}{|\bbox{k}|}$, and
the integration is over the direction of the unit vector
$\bbox{V}$. Some further algebra then leads directly to
 \begin{eqnarray}
 \label{Rp}
   \Sigma_R(p) = \frac{m_\beta^2}{4\pi}\int d\Omega \, 
   \frac{{\not\! V}_+}{V_+\cdot p+i\epsilon}  \, .
 \end{eqnarray}
The advanced electron self-energy in HTL approximation is computed
similarly as
 \begin{eqnarray}
 \label{A}
   \Sigma_A(p) = \frac{m_\beta^2}{4\pi}\int d\Omega\, 
   \frac{{\not\! V}_+}{ V_+\cdot p'-i\epsilon} \, .
 \end{eqnarray}
From this one obtains
 \begin{mathletters}
 \label{46}
 \begin{eqnarray}
 \label{46a}
   \Sigma_R(p')-\Sigma_R(p) &=& 
   - \frac{m_\beta^2}{4\pi}\int d\Omega\,
    \frac{V_+\cdot(p'-p)\, {\not\! V}_+}
         {(V_+\cdot p+i\epsilon)(V_+\cdot p'+i\epsilon)}\, ,
 \\
 \label{46b}
   \Sigma_R(p')-\Sigma_A(p) &=& 
   - \frac{m_\beta^2}{4\pi} \int d\Omega\,
   \frac{V_+\cdot(p'-p){\not\! V}_+}
        {(V_+\cdot p-i\epsilon)(V_+\cdot p'+i\epsilon)} \, .
 \\
 \label{46c}
   \Sigma_A(p')-\Sigma_A(p) &=& 
   - \frac{m_\beta^2}{4\pi} \int d\Omega\,
   \frac{V_+\cdot(p'-p){\not\! V}_+}
        {(V_+\cdot p-i\epsilon)(V_+\cdot p'-i\epsilon)} \, .
 \end{eqnarray}
 \end{mathletters}
If we set in (\ref{12}) $k_1=p, k_2=-p', k_3=p'-p$ and then compare it
with Eqs.~(\ref{46}) we find the following relations
between the hard thermal loop contributions to the 2- and 3-point
functions:
 \begin{mathletters}
 \label{13}
 \begin{eqnarray}
 \label{13a}
   (p'-p)_\mu G_{Ri}^\mu(p,-p',p'-p) &=&
   - i g [\Sigma_R(p')-\Sigma_R(p)]\, ,
 \\
 \label{13b}
   (p'-p)_\mu G_{R}^\mu(p,-p',p'-p) &=&
   - i g [\Sigma_A(p')-\Sigma_A(p)]\, ,
 \\
 \label{13c}
   (p'-p)^\mu G_{Ro}^{\mu *}(p,-p',p'-p) &=&
   - i g [\Sigma_A(p')-\Sigma_R(p)]\, .
 \end{eqnarray}
 \end{mathletters}
These relations are structurally similar to the zero temperature 
Ward identity (\ref{wdT1}) and agree (up to notational differences)
with the general  finite temperature Ward identities in RTF given in
Eq.~(3.17) of Ref.~\cite{olivo97}.

\section{Conclusions}
\label{sec5}

We studied the QED 3-point vertex function at finite temperature
in the CTP real-time formalism. This formalism has recently gained
increased popularity because it allows for a generalization to
non-equilibrium situations as encountered, e.g., in the initial stages
of heavy-ion collisions, and it avoids the need for analytical 
continuation which plagues the imaginary time formalism. We started by
giving a set of useful relations among the eight thermal components of
the real-time vertex function. We then derived spectral integral
representations for the three retarded 3-point functions and
calculated the corresponding spectral densities explicitly at 1-loop
order in the HTL approximation. In this approximation the two
independent spectral densities become degenerate and are transverse
with respect to all three external momenta. 

Inserting these HTL spectral densities into the spectral
representation we obtained three retarded 3-point vertex functions,
two of which turned out to be identical to the corresponding
QCD-analogues derived in \cite{Blz} from a set of classical kinetic
equations in the long-wavelength limit. By contracting the three
3-point vertex functions with the momentum vector of the photon, we
obtained a result which we could compare with the fermion HTL self
energy. The result was a set of real-time Ward-Takahashi identities at
finite temperature which generalize the zero temperature Ward identity
and agree with the recently derived finite temperature identities of
Ref.~\cite{olivo97}.  

Due to the matrix structure of the real-time thermal Green functions, 
there is a whole class of finite temperature Ward identities which
relate retarded and advanced vertex functions to combinations of
retarded and advanced fermion self energies. As first observed by
D'Olivo {\em et al.} \cite{olivo97}, if the ingoing (outgoing) fermion
leg has the largest time, the Ward identity involves only the retarded
(advanced) fermion self energies; if the photon leg has the largest
time, both retarded and advanced fermion self energies are involved.

Clearly, no fundamentally new physical results were derived in this
paper; this was not our goal. What we have achieved is a consistent
real-time representation of the 2- and 3-point functions in finite
temperature field theory, taking into account the full matrix
structure arising from the doubling of degrees of freedom, which makes
their analytic structure explicit and can thus serve as a basis for
non-perturbative resummation schemes at finite temperature. We have
tested the consistency of the formalism in the context of HTL
resummation within thermal equilibrium QED. We expect the formalism to
provide a stable basis for an extension to the real-time dynamics of
non-equilibrium systems. 

\acknowledgments
This work was supported by the Deutsche Forschungsgemeinschaft (DFG),
the Bundesministerium f\"ur Bildung und Forschung (BMBF), the National
Natural Science Foundation of China (NSFC), and the Gesellschaft f\"ur
Schwerionenforschung (GSI). 


\end{document}